\documentclass{article}
\usepackage{graphicx}
\def\espacefig{\bigskip}
\def\real{{{\rm I}\!{\rm R}}}
\def\complex{{\,\,\vrule depth0pt  \!\!{\rm C}}}

\newcommand\maath{\mathsurround=0pt}
\newcommand{\EQM}[1]{\vcenter{\normalbaselines\maath
    \ialign{${\displaystyle ##}$\hfil&&\ ${\displaystyle ##}$\hfil\crcr
    \mathstrut\crcr\noalign{\kern-\baselineskip}
    \noalign{\smallskip}
    #1\crcr\mathstrut\crcr\noalign{\kern-\baselineskip}}}}

\begin{document}

\centerline{\bf Euler configurations and quasi-polynomial systems}
\bigskip
\centerline {Alain Albouy, albouy@imcce.fr}
\centerline {CNRS-UMR 8028, Observatoire de Paris}
\centerline {77, avenue Denfert-Rochereau}
\centerline {75014 Paris, France}
\bigskip
\centerline{Yanning Fu, fyn@pmo.ac.cn}
\centerline{Purple Mountain Observatory}
\centerline{2 West Beijing Road}
\centerline{Nanjing 210008, P.\ R.\ China}
\bigskip
\centerline{3/2006}

\bigskip\bigskip\bigskip

{\sl Abstract.} In the Newtonian 3-body problem, for any choice of the three masses, there are exactly three Euler configurations (also known as the three Euler points). In Helmholtz' problem of 3 point vortices in the plane, there
are at most three collinear relative equilibria. The ``at most three" part is common to both statements, but the respective arguments for it are usually so different that one could think of a casual coincidence. By proving
a statement on a quasi-polynomial system, we show that the ``at most three" holds in a general context which includes both cases. We indicate some hard conjectures about the configurations of relative equilibrium and suggest
they could be attacked within the quasi-polynomial framework.

\bigskip

\centerline{\bf 1. Introduction}

Statics is the science that studies the equilibria of a mechanical system. The notion of {\sl
relative equilibrium} generalizes the notion of equilibrium in the case of a mechanical system with a
continuous symmetry group. This group is often a group of rotations. In a motion
of relative equilibrium the configuration is not required to be fixed, as for an equilibrium. But it is
fixed ``up to symmetry", i.e.\ often ``up to rotation".

The two examples we start with are basic and well-known. In 1762, Euler [Eu1] considered the Sun,
the Earth and the Moon as three point particles moving under Newtonian gravitation.
He discovered a type of motion where the three bodies permanently form a collinear configuration. Each
particle describes an elliptic trajectory. The trajectories may also be hyperbolic, parabolic, circular or rectilinear.
In the circular case, we have a relative equilibrium of the 3-body
problem. In one of the possibilities described by Euler, the Moon
is four times farther from the Earth than it is today, and from
the Earth it is seen as a permanent full moon. Many philosophical
debates about the perfection of the world were raised by Euler's
permanent full moon. They were closed by Liouville who proved that
the relative equilibrium is unstable [Lut].

The second example is the problem of three Helmholtz' vortices. We consider a perfect fluid with an
infinite horizontal surface and a constant thickness. Some states of the fluid are completely described
giving the positions and the vorticities of a finite number of ``point vortices". These vortices are
centered on a vertical line and are sometimes called line vortices. We forget the vertical direction
and just describe the vortices as points in the horizontal plane.
Helmholtz found the ordinary differential system modeling the motion of the vortices. There exist
relative equilibria, i.e.\ configurations that remain unchanged up to rotation and translation. This is
possible with a collinear configuration of $N$ vortices.

Some famous experiments by A.M.\ Mayer simulate Helmholtz equations
by a device where $N$ identical magnets are floating on a surface of water.
This reminds us that equations similar to Helmholtz' are quite
frequent in physical models. Mayer found several equilibrium
configurations. He did not detect any of the collinear equilibria, which
are unstable for any $N\geq 3$, as are the collinear equilibria of $N$ equal vortices
(see [Are] and [ANS]).

In each example we want to determine the set of relative equilibria. To do so we have
to solve a system of equations. Here as in other situations the system may be reduced to a polynomial
system, and we are interested in the real solutions.

{\sl Real solutions of equations.} There are many histories of the theory of equations. The most popular
ones relate how blind were our ancestors who in the 15th century ignored the negative roots of a
polynomial equation. They continue telling the discovery of complex numbers, the fundamental theorem of
algebra, and the Galois group of a polynomial equation with rational coefficients. This is actually the
history of how the initial questions were forgotten.

The initial questions were about the real solutions, often the positive ones. This is true for the
question of determining the equilibria  of a mechanical system. It is also true for the geometrical,
optical or accountancy problems which since the antiquity have motivated the study of complicated
equations.

If {\sl some} traditions did forget the real questions, others
studied them carefully. A central achievement is Descartes' rule of
signs. It bounds the number of positive solutions of a polynomial
equation $\alpha_0+\alpha_1x+\cdots+\alpha_n x^n=0$ by the number of
sign variations in the list $\alpha_0$, $\alpha_1,\dots,\alpha_n$ of
its real coefficients. In his famous works on equations, Lagrange
introduced the reasoning on the permutation of roots which lead to
Galois theory. But he also developed completely distinct ideas,
presenting methods to get the exact number of real solutions if the
coefficients are given numbers. In [Lagr], he emphasizes that he
does not know how to discuss the number of real solutions if some
coefficients of the equation depend on parameters.

Important advances are due to 19th century mathematicians. Fourier
noticed that if we replace Descartes' list of coefficients with the
``list of successive derivatives at a point", we get an upper bound
for the number of roots in any given interval. Vincent showed that a
clever use of rational transformations allows to decrease the number
of sign variations. Sturm changed Fourier's list into the ``list of
values at a point of the successive remainders in the Euclidean
algorithm applied to the polynomial and its derivative", and got the
exact number of roots in an interval. Hermite found a quadratic form
the signature of which is the number of real roots of the given
equation. However, despite many recent continuations of these
classical works, the discussion of the number of real roots when
some parameters vary remains difficult.

Laguerre [Lagu] developed Descartes' rule in another direction.
We consider the expression
$$P(x)=\alpha_1x^{\beta_1}+\cdots+\alpha_n x^{\beta_n}, \quad\hbox {with}\quad
(\alpha_1,\dots,\beta_n)\in\real^{2n}.\eqno (1.1)$$
Laguerre proved that the number of positive roots of the equation $P(x)=0$ is not greater than
the number of sign variations in the list
$\alpha_1,\dots,\alpha_n$, assuming that the monomial terms of $P$
are ordered in such a way that $\beta_1<\cdots<\beta_n$. Of course
we ignore the null $\alpha_i$'s in the count of sign variations.
The proof is: let $i$ be the first index such that the sign of
$\alpha_i$ is not the sign of $\alpha_1$. Consider
$Q(x)=\bigl(x^{-\beta_i}P(x)\bigr)'$. Count the sign variations in
the list $\alpha_1(\beta_1-\beta_i),\dots,\alpha_n(\beta_n-\beta_i)$ of the
coefficients of $Q$. Compare to $P$: there is one sign variation
less. Use a recurrence hypothesis on the number of sign
variations, apply Rolle's theorem and conclude.

Laguerre's statement includes Descartes' rule of signs and is just as simple. It leads us out of the
world of algebraic equations. We accept irrational exponents, even if this seems useless in the
applications. The interesting new feature is that the exponents may be varied continuously. The natural
observation that the number of real roots remains bounded while the exponents are varied is now
understandable.

Our relative equilibria are given by an equation where an exponent called $b$ varies. When $b=-1$ the
equation is algebraic of degree 3 and defines relative equilibria of vortices. When
$b=-2$ it has degree 5 and defines central configurations of celestial bodies. Simple and sharp
statements about the number of real solutions in these problems belong to the theory of Laguerre's type
systems. This is also the theory of quasi-polynomial systems, as we will explain in \S 6.

\bigskip

\centerline{\bf 2. Euler Configurations}

\noindent 2.1. {\sl  The equations of motion.} Let $x_i$ and $m_i$ be respectively the abscissa and the
{\sl mass} of the {\sl particle} $i$, and call $(x_1,\dots,x_n)\in\real^n$ a {\sl collinear
configuration}. We assume
$(m_1,\dots,m_n,b)\in\real^{n+1}$ and set
$$\gamma_i=\sum_{k\neq i} m_k\rho(x_{ki}),\qquad x_{ki}=x_i-x_k,\qquad
\rho(x)=x|x|^{b-1}.\eqno(2.1)$$
If $b=-2$,  $x_{ij}\neq 0$ for any $i$, $j$, $1\leq i<
j\leq n$, and
$m_i>0$ for any
$i$, the Newtonian equations for the particles on the line are $$\ddot x_i=-\gamma_i.\eqno(2.2)$$ There
is a second physical interpretation: we consider the
$n$ collinear particles as Helmholtz' vortices in the Euclidean plane, with vorticities
$m_i\in\real$, and ordinates $y_i=0$. Then Helmholtz' law is $\dot x_i=0$, $\dot y_i=\gamma_i$
where $\gamma_i$ is given by Formula $(2.1)$ with
$b=-1$. We assume
$x_{ij}\neq 0$ for any $i$, $j$,
$1\leq i< j\leq n$.

\noindent 2.2. {\sl Relative equilibria. Central configurations.
Euler's and Moulton's configurations.} The collinear central configurations are, by
definition, the collinear configurations
$(x_1,\dots, x_n)$ such that there exists a
$\lambda\in\real$ with
$$\gamma_j-\gamma_i=\lambda x_{ij}\qquad 1\leq i< j\leq n.$$
They are also called {\sl Moulton configurations}, and in the case $n=3$, {\sl Euler
configurations}. These terminologies come from Celestial Mechanics.

In the Newtonian $n$-body problem, if we associate to an initial
Moulton configuration $(x_1,\dots,x_n)$ the initial velocities
$(\dot x_1,\dots,\dot x_n)=\nu(x_1,\dots,x_n)$, where $\nu\in\real$,
the motion is homothetic. If we associate to the configuration in a
plane $\bigl((x_1,0),\dots,(x_n,0)\bigr)$ the velocities
$\sqrt{\lambda}\bigl((0,x_1),\dots,(0,x_n)\bigr)$, the motion is of
relative equilibrium. In the Helmholtz problem, where $b=-1$, a
collinear central configuration of vortices has a motion of relative
equilibrium: during the motion, the distances between particles
remain constant.

The conditions for central configuration express that the
$n$-uple $(x_1,\dots, x_n)$ and the $n$-uple $(\gamma_1,\dots\gamma_n)$ are equal up to a
translation and a scaling.

This may be written as
$$P_{ijk}(x_1,\dots,x_n)=\left|\matrix{1&1&1\cr
x_i&x_j&x_k\cr\gamma_i&\gamma_j&\gamma_k}\right|=0,\qquad 1\leq
i<j<k\leq n.\eqno(2.3)$$
Translating or rescaling a configuration solution of $(2.3)$ we obtain other solutions. If $n=3$ we normalize the
configurations by putting $x_1=0$, $x_2=1$, $x_3=1+s$. The
equation for normalized Euler's configurations is the single
equation $(2.3)$, written as
$$g(s)=P_{123}(0,1,1+s)=0.\eqno(2.4)$$
Euler stated and proved the existence and uniqueness of a central
configuration of three particles of given positive masses and given ordering. Here is the statement in our
notation.

2.3. {\sl Proposition.} If $m_i>0$, $i=1,2,3$, Equation  $(2.4)$ with $b=-2$ has one and only one
positive solution.

The proof appears in
\S 8 of [Eu2]. Assuming $s>0$, we get
$$\EQM{&\scriptstyle{(1+s)^2s^2g(s)=m_1s^2(1-(1+s)^3)+m_2(1+s)^2(1-s^3)+m_3((1+s)^3
-s^3)}\cr
&\scriptstyle{=-(m_1+m_2)s^5-(3m_1+2m_2)s^4-(3m_1+m_2)s^3+(m_2+3m_3)s^2+(2m_2+3m_3)s+m_2+m_3.}
}\eqno(2.5)$$
Euler writes ``eumque unicum elici, cum
unica signorum variatio occurat''.
The sequence of coefficients changes sign exactly once. By Descartes' rule of signs
there is exactly one positive root. The success of this argument is surprising: it is not
often that Descartes' rule gives an answer for all the required values of the
parameters. Moreover, we check that  the
argument works for any negative integer value of $b$. In particular we put $b=-1$ and find
$$\EQM{&\scriptstyle{(1+s)sg(s)=m_1s(1-(1+s)^2)+m_2(1+s)(1-s^2)+m_3((1+s)^2
-s^2)}\cr
&\scriptstyle{=-(m_1+m_2)s^3-(2m_1+m_2)s^2+(m_2+2m_3)s+m_2+m_3.}
}\eqno(2.6)$$
We notice that for the odd integer values of $b$ the
function $\rho$ in $(2.1)$ is simplified as $\rho(x)=x^b$.
Contrary to $(2.5)$, the rational expression $(2.6)$ of $g(s)$ is
valid for $s<0$. We get this other result.

2.4. {\sl Proposition.} For any given $(m_1,m_2,m_3)\in\real^3\setminus\{(0,0,0)\}$,
$(2.4)$ with $b=-1$ has at most three roots on its domain of definition
$\real\setminus\{-1,0\}$.

In less precise words, there are at most three central configurations of three collinear vortices, for
given vorticities but free ordering of the vortices (in our favorite language given masses but
free ordering of the particles). Note that we are always thinking of ``distinguishable particles".
Counting configurations of indistinguishable particles gives more complicated results (see [LoS]).

In both cases $b=-1$ and $-2$, Euler's argument gives {\sl exactly} three central
configurations, one for each ordering, if $m_i>0$, $i=1,2,3$.
If the signs of the $m_i$ are arbitrary, we use the argument of the degree. For $b=-1$ this gives
Proposition 2.4. For $b=-2$ it gives 15 as the upper bound for the number of central configurations:
there is a different polynomial
$(2.5)$ of degree 5 for each of the 3 orderings. These crude arguments give the upper bounds:
$$\matrix{b&-1&-2&-3&-4\cr \hbox{bound } m_i\in\real&3&15&7&27\cr\hbox{bound }m_i>0&3&3&3&3}\eqno(2.7)$$
The first bound is from the most elementary {\sl algebraic
geometry}. It uses the upper bound on the number of roots given by the degree. The second bound is from
the most elementary {\sl real algebraic geometry}. It uses Descartes' rule of signs. The word algebraic
indicates that we consider polynomial equations, and treat them staying in the framework of polynomial
techniques.

The bound 15 in Table $(2.7)$ is far from optimal. We ask the question:

2.5. {\sl Question.} Fix $b=-2$ and consider all the possible values of
the parameter $(m_1,m_2,m_3)\in\real^3\setminus \{(0,0,0)\}$. What is the maximal number of
roots of Equation $(2.4)$ in the domain $\real\setminus\{-1,0\}$?

To answer this question, we found better to forget the polynomial expansion $(2.5)$ and to prove that 3
is the maximum number of roots for any {\sl real} negative $b$. We use the most elementary {\sl real
analysis}, based on Rolle's theorem. Paradoxically this simplifies the quest for a
proof. For non-integer exponents there are less  available techniques, and the successful ones are among
them. We need less attempts to determine which is the technique that works.

\bigskip
\centerline{\bf 3. Generalized Euler's configurations. The results.}

We describe the set of Euler configurations, i.e.\ collinear central
configurations of three particles, under the general hypothesis
$(m_1,m_2,m_3,b)\in\real^4$.

3.1. {\sl The three cells.} There are 6 possible orderings of the
particles, and the number becomes 3 after identifying an ordering
with its reversed ordering. We say that there are 3 {\sl cells}, one
for each pair of orderings, in the space of collinear
configurations. Each cell, i.e.\ each pair of orderings, has the
name of the particle in the middle.

For example the second cell corresponds to the orderings
$x_1<x_2<x_3$ and $x_3<x_2<x_1$. We always consider strict
orderings. {\sl By convention the configurations with collision are
not central configurations.}

\espacefig \centerline{\includegraphics [width=110mm]
{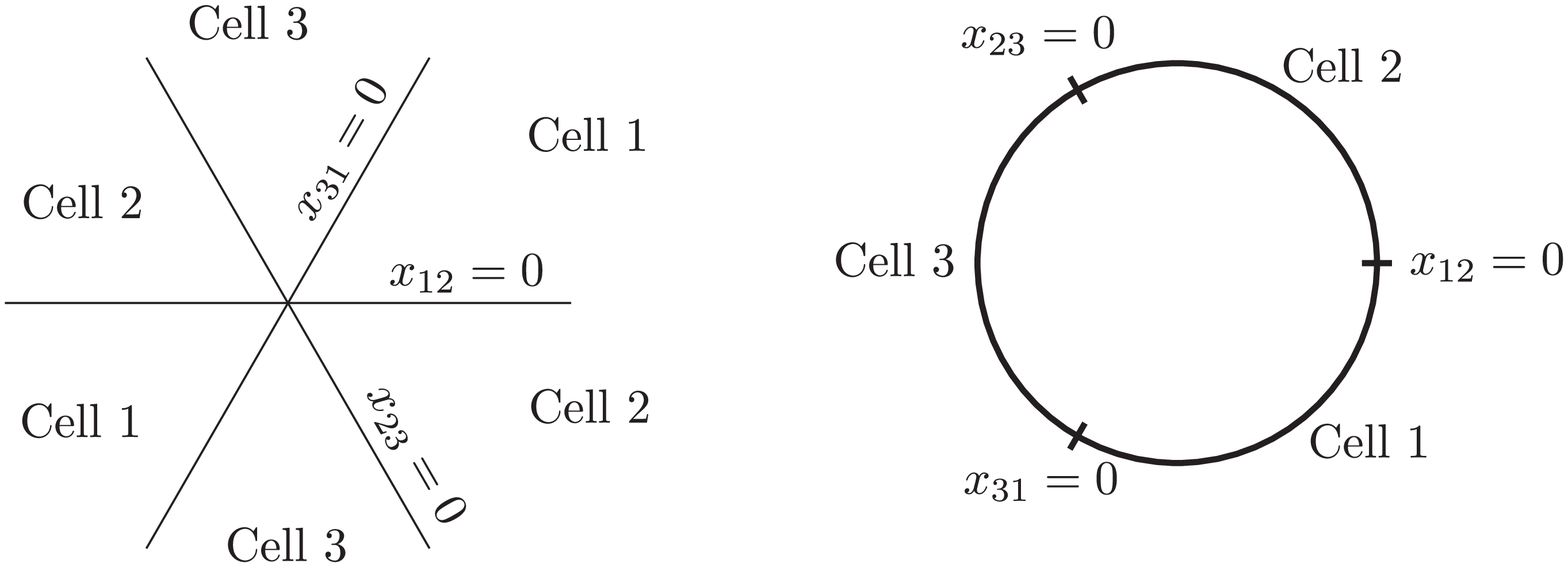}} \nobreak 3.2. {\sl Figure}

We already used the parameterization $(x_1,x_2,x_3)=(0,1,1+s)$.
Later we will also use the parameterization $(x_1,x_2,x_3)=(0,
1-y,1)$. In these parameterizations the cells are indicated below.

\espacefig \centerline{\includegraphics [width=100mm]
{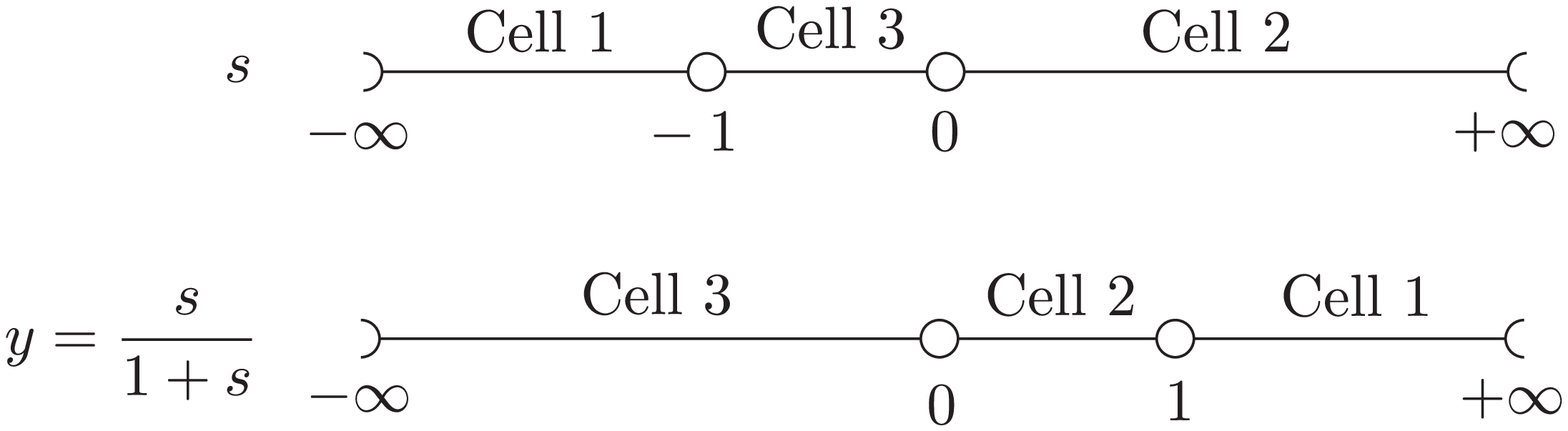}} \nobreak 3.3. {\sl Figure}

\noindent 3.4. {\sl Definition.} Let $g$ be the function of $s$
defined at $(2.4)$. We denote by ${\cal E}_i$ the number of zeros of
$g$ which are strictly inside Cell $i$. We put ${\cal E}={\cal
E}_1+{\cal E}_2+{\cal E}_3$.

Thus ${\cal E}_i$ is the number of ``classes'' of Euler
configurations with particle $i$ in the middle (two configurations
are in the same class if they are homothetic, with a positive or a
negative factor.) We present bounds on ${\cal E}_i$ and then on
${\cal E}$. If there is no restriction on the masses, the situation
is the same for all three cells. We only describe the second cell.

\noindent 3.5. {\sl Proposition.} Let $m_1$, $m_2$, $m_3$ be
arbitrary real masses. The function $g$ vanishes identically on
$]0,+\infty[$ only in the following cases: (i) $m_1=m_2=m_3=0$, (ii)
$b=0$ and $m_1=-m_2=m_3$, (iii) $b=1$, (iv) $b=2$, $m_2=0$ and
$m_1=m_3$, (v) $b=3$ and $m_1=m_2=m_3$.

\noindent 3.6. {\sl Theorem.} For any
$(m_1,m_2,m_3,b)\in\real^4$, we have ${\cal E}_2\leq 3$, except for
the  $(m_1,m_2,m_3,b)$ characterized in Proposition 3.5 for which
${\cal E}_2=\infty$.

We continue discussing the value of ${\cal E}={\cal E}_1+{\cal
E}_2+{\cal E}_3$. Our main applications being $b=-2$ and $b=-1$, we
decided to restrict the study to the case $b<1$.

\noindent 3.7. {\sl Lemma.} Suppose
$(m_1,m_2,m_3)\in\real^3\setminus\{(0,0,0)\}$ and $b<1$. If
$m_1m_3\leq 0$, ${\cal E}_2\leq 1$.  If $m_1\geq 0$, $m_2\geq 0$,
$m_3\geq 0$, $m_1+m_2>0$, $m_2+m_3>0$ then ${\cal E}_2=1$.
If ${\cal E}_2\geq 2$, there is no zero mass, the exterior masses $m_1$
and $m_3$ have the same sign, and the central mass $m_2$ has the
opposite sign. If ${\cal E}_2\geq 2$ and $b<0$ then moreover
$\inf(|m_1|,|m_3|)<|m_2|$.

\noindent 3.8. {\sl Theorem.} Suppose
$(m_1,m_2,m_3)\in\real^3\setminus\{(0,0,0)\}$. If $0<b<1$, ${\cal
E}\leq 5$. If $b<0$,  ${\cal E}\leq 3$.

\noindent 3.9. {\sl Theorem.} If $b<1$, $m_1>0$, $m_2>0$, $m_3\geq
0$,  ${\cal E}_1={\cal E}_2={\cal E}_3=1$, ${\cal E}=3$.

\noindent 3.10. {\sl Proposition.} Suppose
$(m_1,m_2,m_3)\in\real^3\setminus\{(0,0,0)\}$ and $b=0$. If
$m_1=-m_2=m_3$, then ${\cal E}_1={\cal E}_3=0$, ${\cal E}_2=\infty$.
If we are not in this case, nor in the similar cases $m_1=m_2=-m_3$,
$m_1=-m_2=-m_3$, ${\cal E}_1\leq 1$, ${\cal E}_2\leq 1$, ${\cal
E}_3\leq 1$.

We present a complete discussion of the ${\cal E}_i$'s and of the
${\cal E}$ in the particular case where two masses are equal. All
the real values of $b$ are considered.

\noindent 3.11. {\sl Proposition.} If $m_1=m_3=1$, the number ${\cal
E}_2$ is given in Figure 3.12. The frontiers between the different
regions are two half-lines starting from $(m_2,b)=(-1,1)$ and the
curve $m_2=(2^b-2b)/(b-1)$. On the frontiers, excluding the three
intersections of the curve with the half-lines, ${\cal E}_2=1$.

\espacefig \centerline{\includegraphics [width=100mm] {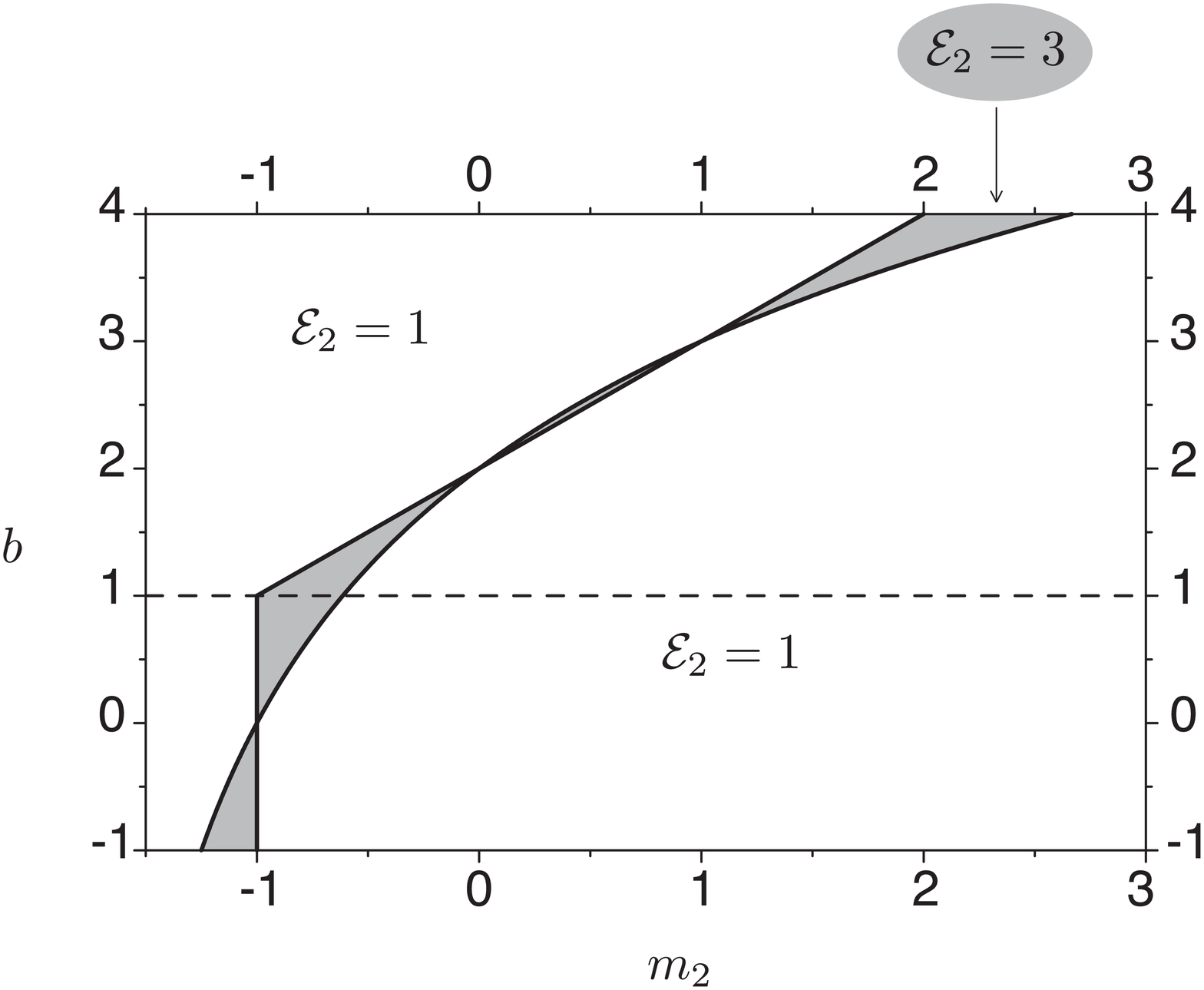}}
\nobreak 3.12. {\sl Figure.} Case $m_1=m_3=1$.

{\sl Remark.} By Proposition 3.5, ${\cal E}_2=\infty$ at the excluded points, as well as on the
line $b=1$. Actually $b=1$ is no longer an exception if we consider $G=g/(b-1)$ instead of $g$. For $b=1$,
$$G(s)=m_1(1+s)\log(1+s)+m_2s\log(s)+m_3(1+s)s(\log(s)-\log(1+s)).$$
If we study $G$, $b=1$ is not a frontier. It is a frontier if we study $g$. We symbolized
this by a dashed line in Figures 3.12 and 3.14.

\noindent 3.13. {\sl Proposition.} If $m_1=m_3=1$, the number ${\cal
E}_1$ is given in Figure 3.14. The frontiers between the different
regions are two half-lines, the same as in Proposition 3.11, and the
upper branch of the hyperbola $m_2(b-1)=2$. On the frontiers,
excluding the intersection of the upper half-line with the
branch of hyperbola, ${\cal E}_1=0$.

In the case $m_1=m_3$ we have ${\cal E}_1={\cal E}_3$. The total
number ${\cal E}={\cal E}_1+{\cal E}_2+{\cal E}_3$ of Euler
configurations is obtained by a mere superposition of Figures 3.12
and 3.14. It gives Figure 3.17. On a frontier ${\cal E}$ is the
minimum of the ${\cal E}$'s of both regions the frontier separates,
if these ${\cal E}$'s are distinct. There are also frontiers for
which ${\cal E}=3$ on both sides. On these frontiers ${\cal E}=1$.

\espacefig \centerline{\includegraphics [width=100mm]
{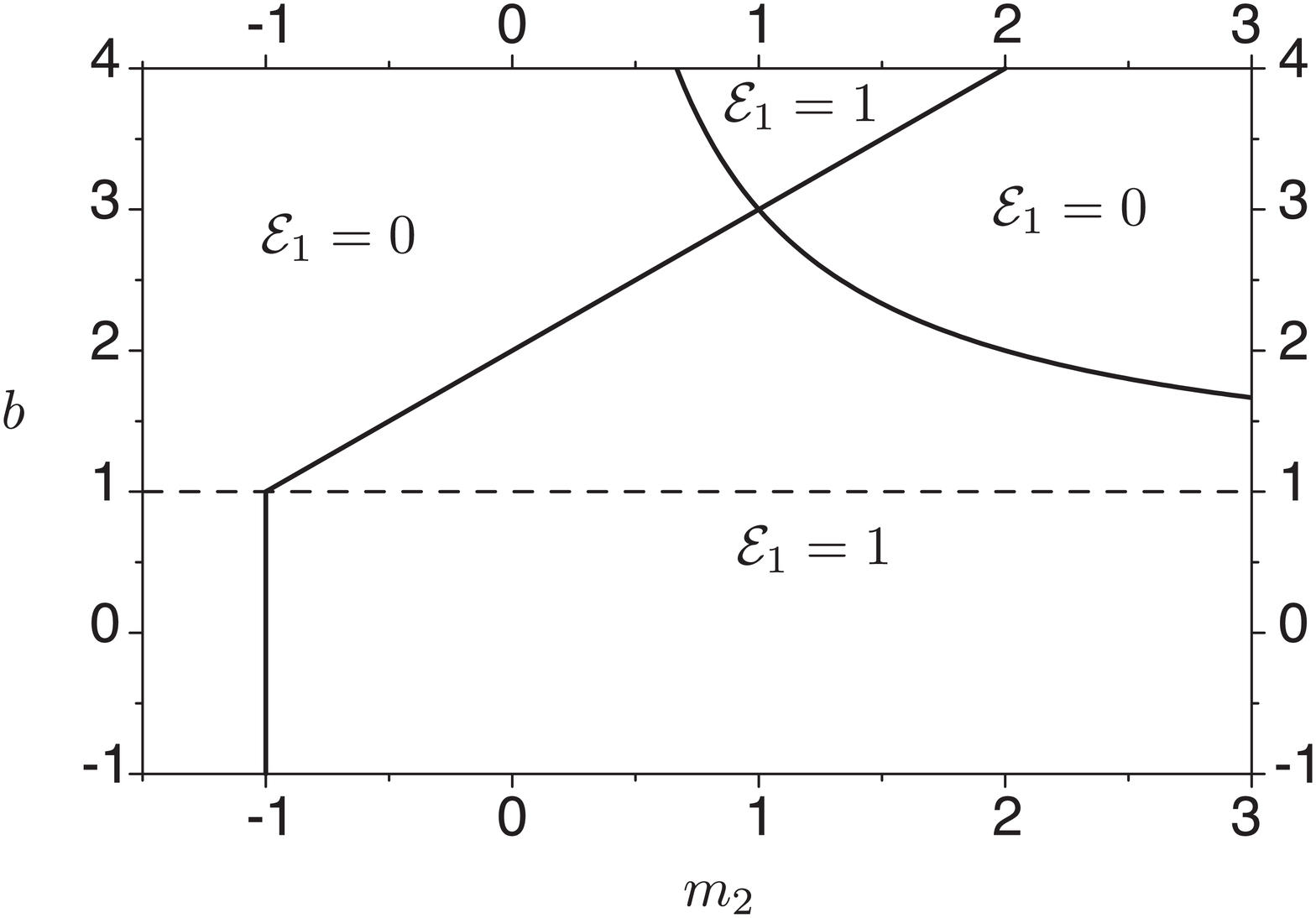}} \nobreak 3.14. {\sl Figure.} Again $m_1=m_3=1$.

3.15. {\sl About lower bounds.} We did not find any lower bound of
${\cal E}_2$ or ${\cal E}$. Figure 3.17 suggests ${\cal E}\geq 1$.
Actually $1$ is a ``generic" lower bound for ${\cal E}$. If we
think of $g$ as a function defined on the circle of Figure 3.2
(the projective line), then its sign changes even number of times.
Since, generically speaking, each of the three collisions between
particles gives a sign change, ${\cal E}$ is an odd number.
However, a collision may be or behave like a double root, and so,
an even ${\cal E}$, $0$ in particular, is not impossible. For
example, if $b \neq 1$ and $(m_1,m_2,m_3)=(0,-1,1)$, {\sl there is
no Euler configuration in any cell}. This may be deduced easily
from the expressions of $A-B$ and $B-C$ we will write in 5.5. Let
us mention an interesting related result (see [Cel]). {\sl For any
Euler configuration with $m_1+m_2+m_3=0$, we have
$m_1x_1+m_2x_2+m_3x_3=0$}. If some mass is non-zero, this
condition fixes the shape of the configuration, which is
non-collisional if and only if $m_1m_2m_3\neq 0$. As a
consequence, for masses satisfying $m_1+m_2+m_3=0$, ${\cal E}=1$
if $m_1m_2m_3\neq 0$, ${\cal E}=0$ if $m_1m_2m_3=0$.

3.16. {\sl About the cases with ${\cal E}=+\infty$.} As Euler
noticed in [Eu2] and [Eu3], the equation of motion $(2.2)$ may be
integrated if the initial configuration is an Euler configuration,
and if the initial velocities are velocities in some homothetic
motion. In the cases where all the configurations are central,
listed in Proposition 3.5, the problem is integrable, for any
initial velocity. This can be easily checked in each case. One can
also notice that if the left hand side of Figure 3.2 is interpreted
as a ``plane of motion", then Equation $(2.2)$ in the cases 3.5
defines a central force problem, and the constant of areas is a
first integral. Case (iii) is a harmonic oscillator. It is
integrated in Proposition 64 of Newton's {\sl Principia}. Case (v)
is discussed by Yoshida [Yos], who looked for all the integrable
cases of the problem of three particles with equal masses on a line,
moving according to $(2.2)$.

\espacefig \centerline{\includegraphics [width=100mm] {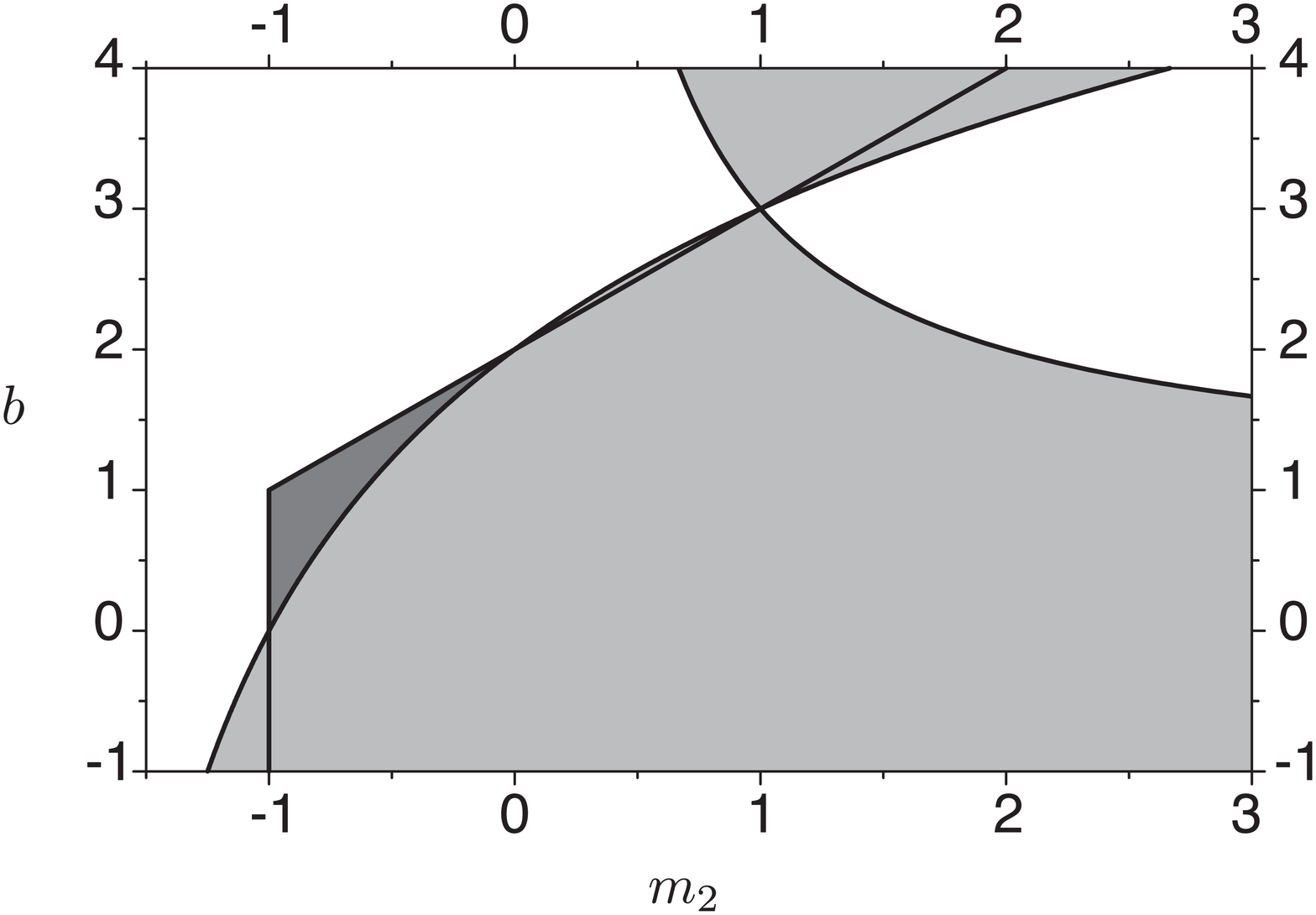}}
\nobreak 3.17. {\sl Figure.} The three orderings together with
$m_1=m_3=1$. White: ${\cal E}=1$. Grey: ${\cal E}=3$. Dark grey:
${\cal E}=5$ (the thinnest region is grey).

\bigskip

\centerline{\bf 4. The formulas}

System $(2.3)$ defines the collinear central configurations. We
compute the term in $m_i$, the notation being again $x_{ij}=x_j-x_i$
and $\rho(x)=x|x|^{b-1}$,
$$m_i\left|\matrix{1&1&1\cr
x_i&x_j&x_k\cr
0&\rho(x_{ij})&\rho(x_{ik})}\right|=m_i\bigl(x_{ij}\rho(x_{ik})-x_{ik}\rho(x_{ij})\bigr)$$
$$=m_ix_{ij}x_{ik}\bigl(|x_{ik}|^{b-1}-|x_{ij}|^{b-1}\bigr).$$
In the case $n=3$, the system is the single equation
$$m_1x_{31}x_{12}\bigl(|x_{12}|^{b-1}-|x_{31}|^{b-1}\bigr)+
m_2x_{12}x_{23}\bigl(|x_{23}|^{b-1}-|x_{12}|^{b-1}\bigr)+$$
$$+m_3x_{23}x_{31}\bigl(|x_{31}|^{b-1}-|x_{23}|^{b-1}\bigr)=0.$$
We normalize the configuration putting $x_1=0$, $x_2=1$, $x_3=1+s$
and arrive at the previous equation $(2.4)$. We make
$$A=(1+s)\bigl(|1+s|^{b-1}-1\bigr),\quad B=s\bigl(|s|^{b-1}-1\bigr),\quad
C=s(1+s)\bigl(|s|^{b-1}-|1+s|^{b-1}\bigr).$$ Equation $(2.4)$ is
$g(s)=0$ with
$$g(s)=m_1A+ m_2B+m_3C.\eqno(4.1)$$
We restrict to the second cell $s>0$ (see Figure 3.3). The following
expansion of $g$ is easy to differentiate several times:
$$g(s)=(m_2+m_3)s^b+(m_1+m_3)(1+s)^b
+m_3\bigl(s^{b+1}-(1+s)^{b+1}\bigr)-m_1(1+s)-m_2s.\eqno(4.2)$$ A
wonderful tool for the study of the equation $g(s)=0$ is the second
derivative $g''(s)$. We get
$$\frac{g''(s)}{b(b-1)}=(m_2+m_3)s^{b-2}+(m_1+m_3)(1+s)^{b-2}
+m_3\frac{b+1}{b-1}\bigl(s^{b-1}-(1+s)^{b-1}\bigr).\eqno(4.3)$$ As
we recalled in the introduction, Laguerre studied the functions of
$y>0$ defined as sums of generalized monomials $\alpha y^\beta$,
where $\alpha$ and $\beta$ are real numbers. There exists a rational
transformation that gives this form to $(4.3)$.
We write:
$$g''(s)=(1-y)^{1-b}H(y), \quad\hbox{with}\quad s=\frac{y}{1-y},\quad\hbox{and}\quad H=b(b-1)h,\eqno(4.4)$$
where
$$h(y)=-\Bigl(m_2-\frac{2m_3}{b-1}\Bigr)y^{b-1}+(m_2+m_3)y^{b-2}-(m_1+m_3){y}+
\Bigl(m_1-\frac{2m_3}{b-1}\Bigr).\eqno(4.5)$$

We have $h(1)=0$. The value $y=1$ corresponds to $s=\infty$ which is
the collision $x_{12}=0$. The correspondence between $y$ and $s$ is
illustrated by Figure 3.3.

\bigskip
\centerline{\bf 5. The proofs}

\noindent{\sl Proof of Proposition 3.5}. We first consider the cases
$b=0,1,2,3$. If $b=0$, $s>0$, $g(s)=m_2+m_3-s(m_1+m_2)$, which is
identically zero if and only if $m_1=-m_2=m_3$. If $b=1$,
$g(s)\equiv0$. If $b=2$, $g(s)=(m_1-m_2-m_3)s+(m_1+m_2-m_3)s^2$,
identically zero if and only if $m_1=m_3$ and $m_2=0$. If $b=3$,
$g(s)=(2m_1-m_2-m_3)s+3(m_1-m_3)s^2+(m_1+m_2-2m_3)s^3$, identically
zero if and only if $m_1=m_2=m_3$. These are the cases (ii) to (v).
For  $b\in\real\setminus\{0,1,2,3\}$, $g\equiv 0$ implies
$b^{-1}(b-1)^{-1}g''\equiv 0$, i.e.\ $h\equiv 0$ for $0<y<1$, and
the four monomials in $(4.5)$ have distinct exponents. The four
coefficients in $(4.5)$ must be zero. This happens only in two
cases. One is $m_1=m_2=m_3=0$ giving the trivial case (i) and the
other $b=-1,m_1=m_2=-m_3\neq 0$ implying $g(s)=m_3(1+2s) \not\equiv
0$. QED.

We present two useful lemmas. They are variations of the various
statements one proves starting from Rolle's theorem. As we use
repeatedly statements as ``the function $y\mapsto \alpha y+\beta$
has at most one root" and do not want to repeat each time ``except
if $(\alpha,\beta)=(0,0)$", we introduce a special way to count the
roots.

5.1. {\sl Definition.} Let $]\mu,\nu[\subset\real$ be an open
interval (bounded or not). If $h:]\mu,\nu[\to\real$ is a function,
we call $Z_{\mu\nu}(h)$ the number of the zeros of $h$ in
$]\mu,\nu[$, with the following exception: if $h\equiv 0$ (i.e.\ $h$
vanishes identically on the interval), $Z_{\mu\nu}(h)=-1$. In the
other cases with an infinite number of roots we simply put
$Z_{\mu\nu}(h)=+\infty$. We say $h$ has a zero at $\mu$ if
$\lim_{x\to \mu} h$ exists and is zero. We say $h$ has a zero at
$\nu$ if $\lim_{x\to \nu} h$ exists and is zero.  We call $\bar
Z_{\mu\nu}(h)$ the number of the zeros of $h$ in $[\mu,\nu]$, with
the same exception: if $h\equiv 0$, $\bar Z_{\mu\nu}(h)=-1$.

5.2. {\sl Lemma.} Let $k\geq 1$ and $m\geq -1$ be two integers.
Let $f:]\mu,\nu[\to\real$ be a $k$ times differentiable function. If
$Z_{\mu\nu}(f^{(k)})\leq m$, then $\bar Z_{\mu\nu}(f)\leq k+m$. If
moreover $\bar Z_{\mu\nu}(f)=k+m$, any root $x\in]\mu,\nu[$ is
non-degenerate, i.e.\ $f'(x)\neq 0$.

5.3. {\sl Lemma.} Consider the expression $P:]0,+\infty[\to\real$,
$y\mapsto P(y)=\sum_{i=1}^n\alpha_iy^{\beta_i}$. For any
$(\alpha_1,\dots,\beta_n)\in\real^{2n}$, $Z_{0\infty}(P)\leq n-1$.
If $n\geq 2$ and $Z_{0\infty}(P)=n-1$, any root is non-degenerate.

{\sl Remark.} This is a simplified version of Laguerre's theorem:
see $(1.1)$. To get the cases with $P\equiv 0$ we simply sum up all
the monomials with same $\beta_i$ and assign zero to all of the resulting coefficients.
We did that in the previous proof.

\noindent{\sl Proof of Theorem 3.6.} The statement to prove is
$Z_{0\infty}(g)\leq 3$. We consider Expression $(4.5)$ of $h(y)$, or rather the corresponding
expression of $H(y)$ which has no denominator.
Lemma 5.3 gives $Z_{0\infty}(H)\leq
3$. As $H(1)=0$, $Z_{01}(H)\leq 2$. By Relation $(4.4)$,
$Z_{0\infty}(g'')=Z_{01}(H)$. By Lemma 5.2, $\bar
Z_{0\infty}(g)\leq 4$. If $b>0$, $g(0)=0$ and $Z_{0\infty}(g)\leq 3$.  In the missing case
$b\leq 0$, let us suppose $Z_{0\infty}(g)=4$.

This is the maximum number allowed, so (i) the four roots of $g$ are non-degenerate, (ii) $b\neq 0$
 (iii) $Z_{0\infty}(g'')=Z_{01}(h)=2$, $Z_{0\infty}(h)=3$, (iv) the three roots of $h$ are
non-degenerate, (v) the four coefficients of $h$ are all non-zero.

By $(4.5)$ and (v), $m_2+m_3\neq 0$ and $m_1+m_3\neq 0$. The sign of
$h$ at zero is the sign of $m_2+m_3$. At $+\infty$ it is the sign of
$-m_1-m_3$. By (iv) this signs are opposite: $(m_2+m_3)(m_1+m_3)>0$.

But  $Z_{0\infty}(g)=4$ means there are four Euler configurations
such that the particle at the left has mass $m_1$, the particle at
the middle has mass $m_2$, the particle at the right has mass $m_3$.
By reflection of the configuration there are also four Euler
configurations such that the particle at the left has mass $m_3$,
the particle at the middle has mass $m_2$, the particle at the right
has mass $m_1$. We can make the deduction above after switching the
indices $1$ and $3$ everywhere. We get the inequality
$(m_2+m_1)(m_1+m_3)>0$.

By $(4.2)$ the sign of $g$ at zero is the sign of $m_2+m_3$. At
$+\infty$ it is the sign of $-m_1-m_2$. According to (i) these signs
coincide. This gives the conflicting inequality
$(m_2+m_3)(m_1+m_2)\leq 0$. QED

5.4. {\sl Non-degenerate roots.} We applied above a quite general
principle: if the number of roots is maximal, the roots are
non-degenerate. The principle is true in the context of
Theorem 3.6: {\sl If ${\cal E}_2=3$, the three roots of $g$ are
non-degenerate.} To prove this, we consider the only opposite case not violating the established assertion
$Z_{0\infty}(g'')\leq 2$: $g$ has a double root and two non-degenerate roots. It implies $Z_{0\infty}(g')\geq
4$ if $b>0$, $Z_{0\infty}(g')\geq 3$ if $b\leq 0$. This is exactly
the same situation as if $g$ had four roots. It can be excluded in the same way.

5.5. {\sl More inequalities.} Expression $(4.1)$ is
$g(s)=m_1A+m_2B+m_3C$. Assuming $b<1$ and $s>0$ we have
$A=(1+s)\bigl((1+s)^{b-1}-1\bigr)<0$, $B=s(s^{b-1}-1)$ and
$C=s(1+s)\bigl(s^{b-1}-(1+s)^{b-1}\bigr)>0$. Besides, as
$$A-B=(1+s)^{b-1}-1+s\bigl((1+s)^{b-1}-s^{b-1}\bigr)<0$$
and as this expression is also $(B-C)/s$, we know that $A<B<C$. On
the other hand, Expression $(4.5)$ may be written as
$h(y)=m_1\alpha+m_2\beta+m_3\gamma$ with $\alpha=1-y>0$,
$\beta=y^{b-2}(1-y)>0$, $\alpha<\beta$ and
$$\gamma=-y+\frac{2}{1-b}(1-y^{b-1})+y^{b-2}>0.$$
In these inequalities we assume $0<y<1$ and again $b<1$. Only the
last one makes problem. To prove it, we compute
$\gamma''(y)=(b-2)y^{b-3}(2y+b-3)>0$. So $\gamma$ is convex on the
interval. Together with $\gamma(1)=0$ and $\gamma'(1)=b-1<0$, this
gives $\gamma>0$.

{\sl Proof of Lemma 3.7.} The first claim is satisfied if moreover
$m_2=0$: $A<0<C$ implies that $g=m_1A+m_3C$ has no root. Therefore, we can prove this claim assuming $m_2 \neq 0$.
Changing if necessary $(m_1,m_2,m_3)$ into $(-m_1,-m_2,-m_3)$ we
may assume $m_2>0$. Switching if necessary the numbering of the
exterior particles $1$ and $3$ we may assume $m_3\ge 0$ and
$m_1\leq 0$. If $m_1\leq -m_2<0\leq m_3$ then
$g=(m_1+m_2)A-m_2(A-B)+m_3C>0$, the second term being positive and the other two terms non-negative
according to the inequalities 5.5. Thus $Z_{0\infty}(g)=0$.  The
other case is $-m_2<m_1\leq 0\leq m_3$. Again by 5.5, we have
$h=(m_1+m_2)\alpha-m_2(\alpha-\beta)+m_3\gamma>0$, which by
Relation $(4.4)$ gives $Z_{0\infty}(g'')\leq 0$. By Lemma 5.2,
$Z_{0\infty}(g)\leq 2$ and if $Z_{0\infty}(g)=2$ both roots are
non-degenerate: $g$ has the same sign at zero and at infinity. But
$g$ has the sign of $m_2+m_3>0$ at zero and the sign of
$-m_1-m_2<0$ at infinity. So $Z_{0\infty}(g)=1$.

The first claim is thus proved, and we know that if ${\cal E}_2>1$
necessarily $m_1m_3>0$. Under the conditions of the second claim, this
means $0<m_1\leq m_3$, forgetting the equivalent $0<m_3\leq m_1$.
Since $m_2\geq 0$, $h=m_1\alpha+m_2\beta+m_3\gamma>0$ by 5.5, and
the signs of $g$ at zero and $+\infty$ are distinct
(implying ${\cal E}_2 \geq 1$). We conclude that $Z_{0\infty}(g)=1$, exactly as we did in the case
$-m_2<m_1\leq 0\leq m_3$.

We now prove the last assertion of Lemma 3.7. We assume ${\cal
E}_2\geq 2$ in the more restrictive case $b<0$. The previous
manipulations on the masses and the proved claims allow us to
assume $m_2<0<m_1\leq m_3$. In order to conclude that $m_1<-m_2$, we
assume $-m_2\leq m_1$ and deduce that ${\cal E}_2=1$.

First we claim that
$$g(s)=m_1(A-B+C)+(m_2+m_1)B+(m_3-m_1)C>0\quad\hbox{if}\quad s\in]0,1[.\eqno(5.1)$$ As $B>0$ if $0<s<1$
and $C>0$ the last two terms are non-negative. We must simply prove
that $A-B+C=2(1+s)^b+s^{b+1}-(1+s)^{b+1}-1>0$. This term is the
expression of $g$ for $(m_1,m_2,m_3)=(1,-1,1)$. If $s=s_0$ is a root
of this expression, $s=1/s_0$ is also a root. As there are at most 3
roots, one being $s=1$, there is at most one root, which is single,
in the interval $]0,1[$. At zero $s^{b+1}$ is the leading term of
$A-B+C$. At $s=1$, the leading term is $(1-s)\Gamma(b)$,
with $\Gamma(b)=2^b-b-1$. As $\Gamma''(b)>0$,
$\Gamma(0)=\Gamma(1)=0$, we have $\Gamma(b)>0$ if $b<0$. Thus
$A-B+C$ is positive at both extremities and consequently in all the
interval $]0,1[$. This proves $(5.1)$. The roots of $g$ are in
$[1,+\infty[$.

We conclude by proving that $g''(s)>0$ if $s\in ]1,+\infty[$. As the
dominating terms of $g$ at $+\infty$ are $-m_1-(m_1+m_2)s<0$, this
means ${\cal E}_2=1$. To prove $g''>0$ we pass to the variable $y$
using $(4.4)$ and prove that $h(y)$ is positive if $y\in ]1/2,1[$.
We have
$$h=m_1(\alpha-\beta+\gamma)+(m_2+m_1)\beta+(m_3-m_1)\gamma.$$
The last two terms are obviously non-negative and we now prove
$\alpha-\beta+\gamma>0$. For this we first note that
$\alpha-\beta+\gamma=1-2y+y^{b-1}+2(1-b)^{-1}(1-y^{b-1})$ is zero at
$y=1$ and positive when $y\to 1^-$. At $y=1/2$ it is
$2^{1-b}(1-b)^{-1}\Gamma(b)>0$, using the notation
$\Gamma(b)=2^b-b-1$ as above. But $\alpha-\beta+\gamma$ is a
Laguerre trinomial vanishing at $y=1$, it cannot vanish more than once or have non-single root
in $]1/2,1[$, so $\alpha-\beta+\gamma>0$ on this interval. QED

5.6. {\sl Table.} In the previous proofs, we used the information contained in the following table
$$\left\{\matrix{&s\to 0\cr b<1&
g(s)=(m_2+m_3)s^b+\varepsilon_0\cr
0<b<1&\varepsilon_0=\bigl((b-1)m_1-m_2-m_3)\bigr) s+o(s^b)\cr b<0&
\varepsilon_0=m_3s^{b+1}+o(s^{b+1})\cr &s\to+\infty\cr
b<1&g(s)=-(m_1+m_2)s+\varepsilon_\infty\cr
0<b<1&\varepsilon_\infty=-\bigl((b-1)m_3-m_2-m_1)\bigr)s^b+o(s^b)\cr
b<0&\varepsilon_\infty=-m_1+o(1) }\right.\eqno(5.2)$$

In the next proof we will need the corresponding information for
other orderings of the particles. While the information can be
obtained by permutation of the indices of the masses, we find the
following description makes it easier to grasp the basic points. As
is obvious, the table describes, for a given ordering, the behavior
of $g$ when two particles tend to coincide. We use subscripts $I$
and $E$ (interior and exterior) to distinguish between the two
particles. The sign of $g$ at the limit is the sign of $m_I+m_E$ or
$-(m_I+m_E)$ according to the collision happens at the right ($s\to
0$, $E=3$ and $I=2$ in the table) or left hand side ($s\to \infty$,
$E=1$ and $I=2$ in the table). If $b<0$ and $m_I+m_E=0$, $g$ has at
the limit the sign of $m_E$ or $-m_E$, again according to the
collision happens at the right or left hand side.

{\sl Proof of Theorem 3.8.} To get ${\cal E}> 3$  we need a cell
with more than one root, thus, by Lemma 3.7, two masses have the
same sign and the third has a different sign, e.g.\ $m_1>0$,
$m_2<0$, $m_3>0$. Cell 2 has at most three roots by Theorem 3.6.
Cells 1 and 3 have at most one each by Lemma 3.7. This gives ${\cal
E}\leq 5$.

If moreover $b<0$, we choose the indices of the masses to get
$0<m_1\leq m_3$. Lemma 3.7 not only gives ${\cal E}_1\leq 1$, but
its proof excludes the possibility of a degenerate root. Consequently
${\cal E}_1=0$ if $(m_1+m_3)(m_1+m_2)<0$, and ${\cal E}_1=1$ if
$(m_1+m_3)(m_1+m_2)>0$. In the same way, ${\cal E}_3=0$ if
$(m_3+m_1)(m_3+m_2)<0$, and ${\cal E}_3=1$ if $(m_3+m_1)(m_2+m_3)>0$.
Our hypothesis ${\cal E}>3$ would imply ${\cal E}_2>1$. By the last
statement of Lemma 3.7, the allowed cases are $0<m_1\leq m_3<-m_2$,
$0<m_1< -m_2< m_3$ and $0<m_1< -m_2=m_3$. In all the cases ${\cal
E}_1=0$. In the first case ${\cal E}_3=0$ and ${\cal E}={\cal
E}_2\leq 3$. In the second case ${\cal E}_3=1$. In Cell 2, $g$ has
the same sign at infinity and at zero. Together with 5.4, this
forbids ${\cal E}_2=3$. So ${\cal E}_2\leq 2$ and ${\cal E}\leq 3$.
In the last case we use 5.6. As $m_2+m_3=0$, we replace $m_2+m_3$ by
the mass of the exterior colliding particle: as $(m_3+m_1)m_2<0$ we get ${\cal E}_3=0$. Again ${\cal E}\leq 3$. QED

{\sl Proof of Theorem 3.9.} See the second statement in Lemma 3.7.
QED

{\sl Proof of Proposition 3.10.} As seen in the proof of 3.5, when
$b=0$, $g(s)=m_2+m_3-(m_1+m_2)s$. A positive root of $g$ gives a
central configuration in Cell 2. For Cell 1 and Cell 3, the
corresponding equations are $m_1+m_3-(m_2+m_1)s=0$ and
$m_3+m_2-(m_1+m_3)s=0$, respectively. The required conclusions
follow immediately.

5.7. {\sl Table.} We will pass to an unrestricted $b$, and in
exchange we will only consider the case of two equal masses. When we
vary the parameters $(m_1,m_2,m_3,b)$ in $\real^4$, the integer
${\cal E}_2$ changes only if  $g$ changes sign at $s=0$ or at
$s=+\infty$ or if  it appears a degenerate root. To determine the
first type of change, we give the expansions corresponding to Table
$(5.2)$ in the case $b>1$.
$$\left\{\matrix{&s\to 0\cr 1< b&
g(s)=\bigl((b-1)m_1-m_2-m_3)\bigr) s+\varepsilon_0\cr 1<b<2&
\varepsilon_0=(m_2+m_3)s^b+o(s^b)\cr 2<b&
\varepsilon_0=b\bigl(m_1(b-1)-2m_3\bigr)s^2/2+o(s^2)\cr&s\to+\infty\cr
1< b& g(s)=-\bigl((b-1)m_3-m_2-m_1)\bigr) s^b+\varepsilon_\infty\cr
1<b<2& \varepsilon_\infty=-(m_2+m_1)s+o(s)\cr 2<b&
\varepsilon_\infty=-b\bigl(m_3(b-1)-2m_1\bigr)s^{b-1}/2+o(s^{b-1})}\right.\eqno(5.3)$$

{\sl Proof of Proposition 3.11.} We set $(m_1,m_2,m_3)=(1,m,1)$,
where $m\in \real$. As the exterior masses are equal, $g(s)=0$ if
and only if $g(1/s)=0$. We have $g(1)=0$, which means that the
symmetric configuration is always a central configuration. By
Theorem 3.6, if $g\not\equiv 0$, there is at most one root $s\in
]0,1[$, which shall be non-degenerate by 5.4. So the determination
of ${\cal E}_2$ is a trivial discussion of the sign of $g$ when
$s\to 0$ and when $s\to 1^-$. The second sign is opposite to that of
$g'(1)=2b-2^b+m(b-1)$. And when $g'(1)=0$, which happens on the curve in Figure 3.12, 1 is a degenerate root.
For the values of $b$ satisfying respectively $b<1$ and $1<b$, the
tables give $g$ at $s=0$: $(1+m)s^b$ and $(b-2-m)s$.
When $(m,b)$ is on the two half-lines, this lowest order term is zero, and the next term
gives the sign of $g$ at zero in the respective intervals $b<0$,
$0<b<1$, $1<b<2$ and $2<b$: respectively $s^{b+1}$, $(b-1)s$,
$(b-1)s^b$, $b(b-3)s^2/2$. A surprisingly simple result appears. In
Figure 3.12, four special points $(-1,0)$, $(-1,1)$, $(0,2)$,
$(1,3)$ divide the connected two half-lines
into five pieces. {\sl On each of these five pieces of line, the
sign of $g$ at zero is the same on the piece of line and on the
white domain it borders.} These being checked, it is easy to finish
the discussion. QED

{\sl Proof of Proposition 3.13.} Instead of setting
$(m_1,m_2,m_3)=(1,m,1)$ and studying the first cell, it is
equivalent to study the second cell setting $(m_1,m_2,m_3)=(m,1,1)$.
This allows to use the tables. For the values of $b$ satisfying respectively $b<1$ and $1<b$, the
tables give $g$ at $s=0$: respectively $2s^b$ and
$\bigl((b-1)m-2\bigr)s$.

For the values of $b$ satisfying
respectively $b<1$ and $1<b$, the tables give  $g$ at $s=+\infty$:
respectively $-(1+m)s$ and $-(b-2-m)s^b$. When $(m,b)$ is on the
two half-lines of Figure 3.14, this leading term is zero. The next term,
respectively $1$, $(1-b)s^b$, $(1-b)s$, $b(b-3)s^{b-1}/2$,
gives the sign of $g$ at infinity in the respective intervals $b<0$,
$0<b<1$, $1<b<2$ and $2<b$. A simple rule can
be observed: on both half-lines the sign of $g$ at
infinity is the sign of $g$ at zero.

\espacefig \centerline{\includegraphics [width=50mm]
{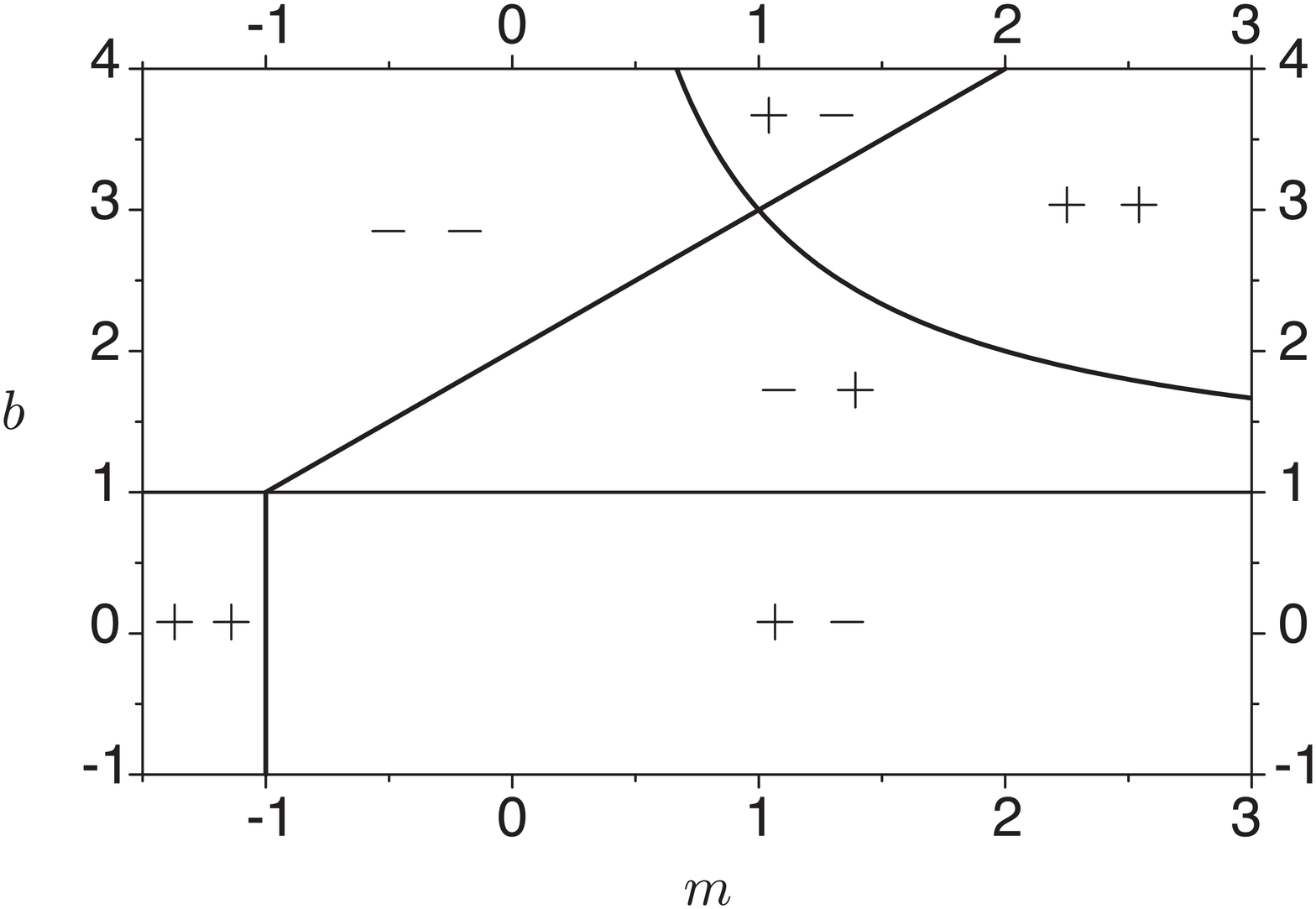}} \nobreak 5.8. {\sl Figure.} Sign of $g$ as
$s\to 0$ and as $s\to +\infty$ for masses $m$, $1$, $1$.

What we need to prove is simply that, while Theorem 3.6 would allow more,
$g$ actually has the least possible number of roots allowed by the signs in the figure.
To treat the part $b<1$, it is sufficient to apply Lemma 3.7: the
middle and the right hand side particles have masses with the same
sign, so there is at most one root.

To treat the case $b>1$, we start with a computation which allows to
improve the general bound ${\cal E}_2\leq 3$ in the case $m_2=m_3=1$.
Here are the expressions $(4.2)$ and $(4.5)$ and the second
derivative of $h$ in this case:
$$g(s)=2s^b+(1+m)(1+s)^b+s^{b+1}-(1+s)^{b+1}-m(1+s)-s,$$
$$h(y)=-\frac{b-3}{b-1}y^{b-1}+2y^{b-2}-(1+m){y}+m-\frac{2}{b-1},$$
$$h''(y)=(b-2)(b-3)y^{b-4}(2-y).$$
As $Z_{01}(h'')\leq 0$, by Lemma 5.2 $\bar Z_{01}(h)\leq 2$. As
$h(1)=0$, $Z_{01}(h)\leq 1$, and $Z_{0\infty}(g'')\leq 1$
according to $(4.4)$. Lemma 5.2 gives $\bar Z_{0\infty}(g)\leq 3$.
In the considered case $b>1$, $g(0)=0$, and so, $Z_{0\infty}(g)\leq 2$.
Except in the case $(m,b)=(1,3)$ which has an infinity of roots, ${\cal E}_2\leq 2$.
If ${\cal E}_2=2$, the roots are non-degenerate.

Let us draw horizontal lines on Figure 5.8. We fix $b>1$ and vary
$m$. There is a ``central interval", empty for $b=3$, which contains
values of $m$ such that $g$ has different signs at zero and at
infinity: on this interval ${\cal E}_2=1$. We focus on both boundary points
of this central interval.

At these points $m=b-2$ or $m=2(b-1)^{-1}$. In the former case, $h'(1)=b-2-m=0$, thus $Z_{01}(h')\leq 0$ due to $Z_{01}(h'')\leq 0$ and to
Lemma 5.2. We continue with the same arguments and get $\bar Z_{01}(h)\leq 1$ and $Z_{01}(h)\leq 0$ because $h(1)=0$.
In the latter case, if $b>2$, $h(0)=m-2(b-1)^{-1}=0=h(1)$.
The bound $\bar Z_{01}(h)\leq 2$ gives $Z_{01}(h)\leq 0$. If $1<b\leq 2$, $h$ is positive as $y\to
0$, and as $y \to 1^-$ because $h'(1)<0$. So $h$ does not vanish in $]0,1[$, since otherwise it would be at a degenerate root or more than
once. Thus $Z_{01}(h)\leq 0$ in all the cases.

Consequently for both boundary points $Z_{0\infty}(g'')\leq 0$ and $Z_{0\infty}(g)\leq 1$ since $g(0)=0$.
If $Z_{0\infty}(g)=1$, the root is non-degenerate. As we noticed, on
the half-line $m=b-2$ the signs of $g$ at zero and at infinity
coincide: $Z_{0\infty}(g)\leq 0$. For the other point
$m=2(b-1)^{-1}$ we have $g'(0)=(b-1)m-2=0$. So $Z_{0\infty}(g')\leq
0$. As $g(0)=0$, $Z_{0\infty}(g)\leq 0$.

So for any $b>1$, $Z_{0\infty}(g)=1$ on the central interval in $m$
and $Z_{0\infty}(g)\leq 0$ on both extremities of this interval. We
observe that the factor of $m$ in the expression of $g$ is
$(1+s)\bigl((1+s)^{b-1}-1\bigr)>0$. Consequently for values of $m$
outside the central interval $g$ is sign definite. QED

\bigskip

\centerline{\bf 6. Quasi-polynomial systems, straight and curved}

6.1. {\sl Quasi-polynomial systems.} Let us write a system in two
variables $x>0$, $y>0$:
$$0=\sum_{i=1}^ma_ix^{b_i}y^{c_i},\quad 0=\sum_{i=1}^n\alpha_ix^{\beta_i}y^{\gamma_i}.\eqno(6.1)$$
The system is polynomial if the exponents  $b_i$, $c_i$, $\beta_i$,
$\gamma_i$ are non-negative integers. However, as in the univariate
case $(1.1)$, we consider that $b_i$, $c_i$, $\beta_i$, $\gamma_i$
may be arbitrary real numbers. This extension fits well with the
restriction $x>0$, $y>0$. What is the correct name for this class of
systems? Some authors call them ``fewnomial systems" (see [Kho]).
But the word ``fewnomial" suggests that the exponents are integers,
and that there are few monomials. Other authors use it with this
more restrictive meaning. Then the word does not indicate that we
can vary the exponents continuously.

Before answering the question of terminology, we should notice that
$(x^a-1)/a$ belongs to the class $(1.1)$. Its limit at $a=0$ is
$\log x$. This suggests to extend the class and accept the $\log$.
In [Ron] the quasi-polynomials are defined as expressions such as
$\sum_{i=1}^ka_ix^{b_i}y^{c_i}$, but after the substitution $x=e^u$,
$y=e^v$. So the quasi-polynomial functions of $(u,v)$ have the form
$\sum a_ie^{b_iu+c_iv}$. The coefficients $a_i$ may be real numbers
as in our favorite cases, but they can also be polynomials in
$(u,v)$. A quasi-polynomial in $(u,v)$ is a polynomial if $k=1$, $b_1=c_1=0$. The function $\log x=u$ is
accepted. In [Kho] the definition of a quasi-polynomial is presented
differently. One chooses a collection of $k$ vectors
$(b_i,c_i)\in\real^2$ and set $E_i=e^{b_iu+c_iv}$. Any polynomial in
$(u,v,E_1,\dots,E_k)$ is a quasi-polynomial. Of course, both
definitions are equivalent. What may differ is the number $k$ of
basic exponential expressions $E_i$.

Equation $(4.2)$ for Euler configurations of Cell 2 (ordering
$x_1<x_2<x_3$) has the type $(6.1)$. It is the system of two
equations in $s>0$, $t>0$,
$$0=s+1-t,$$
$$0=(m_2+m_3)s^b+(m_1+m_3)t^b+m_3\bigl(s^{b+1}-t^{b+1}\bigr)-m_1t-m_2s.$$
Theorem 3.6 gives the optimal upper bound for the number of
solutions of this system, assuming that this number is finite, while
Proposition 3.5 describes the cases where this number is not finite.

Khovanskii obtained general upper bounds for the number of isolated
roots of a quasi-polynomial system. It is well-known that these
bounds are far from optimal, especially when applied to a particular
system as ours. For a system of two equations $P_1=P_2=0$ in two
unknowns $(u,v)$, let $k$ be the number of basic exponential
expressions $E_i=e^{b_iu+c_iv}$. Let $d_i$, $i=1,2$, be the degree
of the polynomial $P_i$ in its variables $(u,v,E_1,\dots E_k)$.
Khovanskii's upper bound is $d_1d_2(d_1+d_2+1)^k2^{k(k-1)/2}$. We
can play to apply this bound to our system. The first point of view
is to consider that there are $k=6$ exponential expressions $s=e^u$,
$t=e^v$, $e^{bu}$, $e^{bv}$, $e^{(b+1)u}$, $e^{(b+1)v}$, and that
the degrees are $d_1=d_2=1$. The bound is $3^62^{15}$. A more
successful point of view is to consider there are only $k=4$
exponential expressions $e^u$, $e^v$, $e^{bu}$, $e^{bv}$, and that
the degrees are $d_1=1$ and $d_2=2$. The bound is $2^{15}=32768$.

Khovanskii's upper bound is obtained considering the polynomial
$P_2$ as a function defined on the curve $P_1=0$. ``Curved''
versions of Rolle's theorem are then used to bound the number of
roots. This process may be adapted to each particular problem, in
order to improve the upper bound. In a recent work, [GNS] obtained
in this way the following result: there are at most 12 equilibrium
points in the Coulombian field of three fixed charges, except if
they are infinitely many. This bound is excellent, but the authors
conjecture that the optimal upper bound is 4 instead of 12.

6.2. {\sl Straight case.} In our case, it is not necessary to create
a curved version of Rolle's theorem. The first quasi-polynomial is
$t=1+s$. It defines a straight segment in the quadrant $s>0$, $t>0$.
The second quasi-polynomial is a function on this segment, and we
can apply the standard Rolle theorem after the explicit substitution
$t=1+s$.

Leandro [Lea] proposes another problem of central configurations
which can be analyzed through the standard Rolle theorem. The
question is the determination of the positions of a test particle
without mass under the attraction of other particles with equal
masses, in such a way that there is relative equilibrium.

When we can reduce the problem to the study of a function on an
interval, we say we are in the ``straight case". A system of two
quasi-polynomial equations is reducible to the straight case if one
of the equations is a trinomial equation
$a_1e^{b_1u+c_1v}+a_2e^{b_2u+c_2v}=a_3e^{b_3u+c_3v}$ with
$(a_1,\dots, c_3)\in\real^9$. As we may always divide the trinomial
equation
 by its right member, we may assume $a_3=1$,
$b_3=c_3=0$. We may set $u'=b_1u+c_1v$, $v'=b_2u+c_2v$, $x=e^{u'}$,
$y=e^{v'}$. The trinomial becomes $a_1x+a_2y=1$ and defines a piece
of line.

A simple process of repeated differentiation with suppression of
some exponential expression at each step leads to the upper bound
$2^n-2$ for the number of roots of a quasi-polynomial systems of two
equations, one being a trinomial and the other a $n$-nomial. This
easy bound is published in [LRW], where is also solved the much
harder problem of the optimal upper bound in the case $n=3$ of two
trinomials. This bound is 5. It is known to be optimal thanks to a
surprising example by Haas of a system of two trinomials with 5
roots. It appeared to be difficult to find examples with more than 4
roots.

The general upper bound $2^n-2$ is 62 in our problem. It is still
far above our optimal upper bound 3. We wish to emphasize that the
main difficulty in finding good upper bounds, in the straight as in
the curved cases, is the research of good variables, good equations,
good repeated derivations, rather than the conceptuality of the
tools. The lack of systematic method is quite a disappointing
observation. Future progresses in this domain could come from an
analysis of the successful choices. We hope that some conclusion may
arrive simplifying or improving our work, [Lea], [LRW] or [GNS].

In the curved case, we do not know any work about the practical
determination of the cases where the number of solutions is
infinite. Probably quite subtle theorems could be used. Many such
theorems are given in [Ron].

\bigskip
\centerline{\bf 7. Open questions on central configurations}

We state a problem of central configurations giving first the number $n$ of particles and the dimension
$p$ of the configuration, which satisfies $1\leq p\leq n-1$. After these integer parameters, we fix
the masses $(m_1,\dots, m_n)\in\real^n$. Finally, we decide the law of attraction: $b=-2$
and $b=-1$ correspond to configurations of vortices and celestial bodies,
respectively. We can also work with any real $b$. We could even consider
non-homogeneous attraction laws, but this would lead us too far away from our preferred applications
$b=-2$ and $b=-1$.

7.1. {\sl Particles on a line.} What is the maximal number of
central configurations with $p=1$ and $n=4$, for a given $b\in\real$
and masses varying in $\real^4$? Of course, we should find together
the answer to the question:
when is this number infinite? We do not know the
answer, even in the case $b=-1$ of the vortices.

The known result is Moulton's theorem, which applies to the case of
positive masses and $b<0$. {\sl There is exactly one central
configuration for each ordering of the masses.} Moulton's theorem is
true for any $n\geq 2$, and gives $n!/2$ central configurations.
Applying this result to the case of $n=3$ particles, we recover one
of the results of Theorem 3.8: ${\cal E}=3$ for $b<0$ and positive
masses. But we proved this result for any $b<1$. Is Moulton's
theorem still true if $0\leq b<1$?

7.2. {\sl Equal masses.} The case $m_1=m_3$ studied in Proposition 3.11 illustrates a general rule: if
the repartition of the mass allows the symmetry of the configurations, and if there are few solutions, then all the solutions are symmetric.
In the case
$n=4$, $p=2$, $b=-1$ or
$b=-2$, and equal masses, it appeared to be possible to prove directly the symmetry (see [Alb]). In contrast, it seems
still impossible, in the case $b=-2$, to get directly a sufficiently low higher bound on the number of solutions, that would
give the symmetry as a consequence.

Consider more generally a central configuration of particles with
equal masses. Does it always possess some symmetry? Numerical experiments by
Moeckel with $b=-2$, $p=2$ showed that the answer is no if
$n=8$, but is probably yes if $n\leq 7$. Numerical experiments
by Kathryn Glass gave the same result for $b=-1$ (see [Gl1], [ANS]). Glass later studied
the bifurcations of the central configurations when $b$ varies, and
found asymmetric configurations with $b=-11$, $p=2$ and $n=6$.
Faug\`ere proved by Gr\"obner base techniques that for
$n\leq 7$, $b=-1$, $p=2$, all the configurations possess some
symmetry. He noticed that each configuration of $n$ indistinguishable
particles $(z_1,\dots,z_n)\in\complex^n$ is associated in a unique way with the polynomial
$P(X)=(X-z_1)\cdots(X-z_n)$. He used the coefficients of $P$ as variables
and found all the central configurations for $n\leq 7$.

7.3. {\sl Four bodies in the plane.} There is an outstanding recent
result for $n=4$, $p=2$, $b=-2$. Hampton and Moeckel proved (see
[HaM]) that there is a finite number of planar central
configurations of 4 particles, whatever be the positive masses. This
solves the first case of an old conjecture by Chazy, repeated by
Wintner and then by Smale in his list of problems for the 21st
century. The finiteness is also conjectured for any $n\geq 5$ and
for $p=2$ or $3$. Indeed all the dimensions $p$ between $2$ and
$n-2$ are interesting and we do not know if there is finiteness when
$n\geq 5$, whatever be the given set of $n$ positive masses. In
contrast, [Rob] shows that there is a continuum of central
configurations if $n=5$, $p=2$, $b\in\real$,
$m_1=m_2=m_3=m_4=-2^{-b}m_5$.

Hampton and Moeckel proved that there are at most 4230 central configurations with $n=4$, $p=2$.
They used Bernstein's ideas. Between the pair ``complex unknowns, positive
integer exponents" and the pair ``positive unknowns, real
exponents", there is the interesting possibility ``non-zero complex
unknowns, integer exponents". Bernstein gave in 1975 a method
to count the solutions with non-zero values of unknowns, and sketched a method to discuss the finiteness. The 4230
configurations are given by the ``mixed volume" method, also used in
[GNS]. They exist as complex solutions. Here we count them
identifying a planar configuration with the reflected configuration.

We know (see [HaM]) that there are at least
$10$ central configurations with $n=4$ and $p=2$, and even
$11$ if we avoid an explicit degenerate case. Numerical experiments indicate $19$ as
the maximum. There are 19 central configurations if
$m_1=m_2=m_3=m_4>0$.

To finish, let us select an unsolved sub-problem: in the case $n=4$,
$p=2$, positive masses, prove that there is exactly one convex
central configuration such that two given particles are not adjacent
(i.e.\ they are on the same diagonal). The existence of such a
convex central configuration is known. ``Convex" means that no
particle is inside the triangle of the three others.
\bigskip

{\sl Acknowledgements.} Fu is supported by NSFC(10473025 and 10233020). We wish to thank Ilias Kotsireas and Carles Sim\'o for precious information.
\bigskip

\centerline{\bf References}

\noindent[Alb] A. Albouy, {\sl The symmetric central configurations of four equal
masses}, Contemporary Mathematics 198 (1996) pp.\ 131--135

\noindent[ANS] H. Aref, P.K. Newton, M.A. Stremler, T. Tokieda, D. Vainchtein, {\sl Vortex Crystals}, Advances in applied
Mechanics 39 (2003) pp.\ 1--79

\noindent[Are] H. Aref, {\sl On the equilibrium and stability of a row of point vortices}, J.\ Fluid Mech.\ 290 (1995)
pp.\ 167--181

\noindent[Cel] M. Celli, {\sl Homographic three-body motions with positive and negative masses},
proceedings of the SPT 2004 conference, Cala Gonone, World Scientific (2004) pp.\ 75--82

\noindent[Eu1] L. Euler,  {\sl Considerationes de motu corporum coelestium}, Novi commentarii academiae
scientiarum Petropolitanae 10 (1764), 1766, pp.\ 544--558 (read at Berlin in april 1762). Also
in Opera Omnia, S.\ 2, vol.\ 25, pp.\ 246-257 with corrections and comments by M. Sch\"urer

\noindent[Eu2] L. Euler, {\sl De motu rectilineo trium corporum se mutuo attrahentium}, Novi
commentarii academiae
scientiarum Petropolitanae 11 (1765)  pp.\ 144--151 (read at St Petersbourg in december 1763). Also in
Opera Omnia S.\ 2, vol.\ 25, pp.\ 281--289

\noindent[Eu3] L. Euler, {\sl Consid\'erations sur le probleme des trois corps}, M\'em.\ de
l'acad.\ d.\ sc.\ de Berlin 19 (1763) pp.\ 194--220

\noindent[Fau] J.C. Faug\`ere, {\sl Probl\`eme des $n$ tourbillons de masses \'egales}, preprint (2000)

\noindent[Gl1] K. Glass, {\sl Equilibrium configurations for a system of N particles in the plane}, Physics Letters A 235 (1997)
pp.\ 591--596

\noindent[Gl2] K. Glass, {\sl Symmetry and bifurcations of planar configurations of the $N$-body and other
problems}, Dynamics and Stability of Systems, 15 (2000) pp.\ 59--73

\noindent[GNS] A. Gabrielov, D. Novikov, B. Shapiro, {\sl Mystery of point charges}, preprint (2005)

\noindent[Haa] B. Haas, {\sl A Simple Counterexample to Kouchnirenko's Conjecture}, Beitr\"a\-ge zur
Algebra und Geometrie, Contributions to Algebra and Geometry, 43 (2002) pp.\ 1--8

\noindent[HaM] M. Hampton, R. Moeckel, {\sl Finiteness of relative equilibria of the four-body problem},
Invent.\ math.\ (2005)

\noindent[Kho] A.G. Khovanskii, {\sl Fewnomials}, Translations of the American Math.\ Society 88
(1991)

\noindent[Lagr] J.L. Lagrange, {\sl R\'eflexions sur la r\'esolution alg\'ebrique des \'equations},
\oe uvres 3 (1770) p.\ 205

\noindent[Lagu] E. Laguerre, {\sl Sur la th\'eorie des \'equations num\'eriques}, Journal de
Math\'e\-matiques pures et appliqu\'ees, s.\ 3, t.\ 9 (1883) pp.\ 99--146. Also in {\sl \OE uvres de
Laguerre, tome 1}, Paris (1898) Chelsea, New-York (1972) pp.\ 3--47

\noindent[Lea] E.G. Leandro, {\sl On the Dziobek configurations of the restricted $(N+1)$-body problem
with equal masses}, preprint (2005)

\noindent[LoS] Y. Long, S. Sun, {\sl Collinear Central Configurations and Singular Surfaces in the Mass
Space}, Arch.\ Rational Mech.\ Anal.\ 173 (2004) pp.\ 151--167

\noindent[LRW] T.-Y. Li, J.M. Rojas, X. Wang, {\sl Counting real connected components of trinomial curve
intersections and $m$-nomial hypersurfaces}, Discrete Comput.\ Geom.\ 30 (2003) pp.\ 379--414

\noindent[Lut] J. L\"utzen, {\sl Joseph Liouville, 1809--1882, master of pure and applied
mathematics}, Sprin\-ger-Verlag (1990) p.\ 653

\noindent[May] A.M. Mayer, {\sl Floating magnets}, Nature 17 \& 18 (1878)
pp.\ 487--488 \& pp.\ 258--260

\noindent[Moe] R. Moeckel, {\sl Some Relative Equilibria of N Equal Masses}, preprint (1989)

\noindent[Mou] F.R. Moulton, {\sl The straight line solutions of the problem of N
bodies}, Ann.\ of Math.\ 2-12 (1910) pp.\ 1--17

\noindent[Rob] G.E. Roberts, {\sl A continuum of relative equilibria in the five-body problem}, Physica D 127 (1999)
pp.\ 141--145

\noindent[Ron] L.I. Ronkin, {\sl Entire functions}, in {\sl Several complex variable III, Geometric
function theory}, Encyclopaedia of mathematical sciences, vol.\ 9, Springer-Verlag (1989) p.\ 26

\noindent[Yos] H. Yoshida, {\sl A criterion for the non-existence of an additional integral in
Hamiltonian systems with a homogeneous potential}, Physica D 29 (1987) pp.\ 128--142

\end{document}